\documentclass[english]{article}
\usepackage[T1]{fontenc}
\usepackage[latin1]{inputenc}
\usepackage{amssymb}

\makeatletter



\usepackage{babel}

\makeatother
\begin{document}
\begin{titlepage}
\thispagestyle{empty}
\hskip 1cm

\begin{center}
{\LARGE{\bf General asymptotic solutions}}\\
\vskip 3mm
{\LARGE{\bf of the Einstein equations and}}\\
\vskip 3mm
{\LARGE {\bf phase transitions in quantum gravity}}\\
\vskip 3mm

\vspace{15pt}

{\large {\bf Dmitry Podolsky}}\footnote{On leave from Landau Institute
for Theoretical Physics, 119940, Moscow, Russia.}

\vspace{10pt}

{Helsinki Institute of Physics, University of Helsinki,}\\
{Gustaf H\"{a}llstr\"{o}min katu 2, FIN00014, Helsinki, Finland}\\

\vspace{10pt}
{Email: dmitry.podolsky@helsinki.fi}\\
\vspace{10pt}
\end{center}

\begin{abstract}
We discuss generic properties of classical and quantum theories of
gravity with a scalar field which are revealed at the vicinity of the
cosmological singularity. When the potential of the scalar field is exponential
and unbounded from below, the general solution of the Einstein equations has
quasi-isotropic asymptotics near the singularity instead of the usual
anisotropic Belinskii - Khalatnikov - Lifshitz (BKL) asymptotics.
Depending on the strength of scalar field potential, there exist two
phases of quantum gravity with scalar field: one with essentially anisotropic
behavior of field correlation functions near the cosmological singularity, and another
with quasi-isotropic behavior. The ``phase transition'' between the two
phases is interpreted as the condensation of gravitons.
\end{abstract}
\end{titlepage}

One pessimistic quotation from the golden era of finding exact solutions of the Einstein equations
which reflected the relations between particle theorists and experts in GR
belongs to Richard Feynman. Taking part in the International
Conference on Relativistic Theories of Gravitation at Warsaw, he was
writing to his wife \cite{Feynman}: ``I am not getting anything
out of the meeting. I am learning nothing. ... I get into arguments
outside the formal sessions (say, at launch) whenever anyone asks
me a question or starts to tell me about his ``work''. The ``work''
is always: (1) completely un-understandable, (2) vague and indefinite,
(3) something correct that is obvious and self-evident but worked
out by a long and difficult analysis, and presented as an important
discovery, or (4) a claim based on the stupidity of the author that
some obvious and correct fact, accepted and checked for years, is
in fact false ... (5) an attempt to do something probably impossible
but certainly of no utility which, it is finally revealed in the end,
fails ... or (6) just plan wrong ... Remind me not to come to any
more gravity conferences!'' Certainly, I am well aware of that the
work presented in this essay could belong to the class (3) or (5) in the
Feynman's classification (hopefully, not to the class (6)!), but I
will follow Feynman's own words \cite{Feynman}: {}``We all do it
for the fun of it'' trying to find my fun in identifying some links
which connect the part of the common lore on general relativity named
{}``Exact solutions of the Einstein equations'' to the problem of the
GR quantization.

Of course, Feynman's interest was in the quantization of GR by applying
the path integral approach working so well in QED. Solutions of the Einstein
equations define saddle points of the action%
\footnote{From now on, by the quantum theory of gravity we mean effective QFT
of spin 2 fields \cite{BurgessEffective} (plus matter fields) ---
the one which particles with energies $E\ll M_{P}$ test. In this
limit, the effects of the non-renormalizability may be neglected.
Although we discuss below the situation which is realized near the
cosmological singularity, we limit the discussion to time scales $t\gg t_{P}$.%
} $S=S_{{\rm gravity}}+S_{{\rm matter}}$ of the quantum gravity with
matter. However, the contributions of these saddle points into the
partition function of the theory and fluctuations near them
\begin{equation}
Z=\int \frac{Dg_{ik}}{Df}D\phi_{{\rm matter}}\exp\left(-\frac{i}{\hbar}\left(S_{{\rm gravity}}+S_{{\rm matter}}\right)\right)
\label{eq:GRPartitionFunctionSmple}
\end{equation}
typically have \emph{zero measure}. In other words, the probability
for an almost any exact solution to describe the observable features of the
Universe or some parts of it, to appear somehow from the quantum foam realized
near the singularity is infinitely small, and the Feynman's anger
is absolutely understandable.

Well, almost absolutely... Of course, there are several classes of solutions which will be important
for the quantum part of the story, too, and one can without much thinking
immediately identify some:

\begin{enumerate}
\item \emph{Attractors}: among them are Minkowski spacetime, de Sitter (at least in the sense of
eternal inflation \cite{LindeEternalInflation}) and anti de Sitter
spacetimes (a set of AdS domains is mostly probably the global attractor
of GR realized as low-energy approximation of string theory \cite{LindeLandscape});
black holes (Schwarzschild, Kerr, Reissner-Nordstr\"{o}m, Kerr-Newman
solutions), etc.
\item \emph{General solutions of the Einstein equations}. As usual
\cite{LL2}, a solution of the Einstein equations is regarded as general
if it contains sufficient number of arbitrary functions of coordinates.
In the case of Ricci-flat spacetimes, this number is $4$, and is equal to $8$ in the presence of hydrodynamic matter.
\end{enumerate}

While any non-attractor type solution of the Einstein equations defines
the saddle point for the path integral (\ref{eq:GRPartitionFunctionSmple}) which does
have a vanishing contribution into the overall partition function,
eventually it well settle down towards an attractor solution due to
the effect of classical perturbations and/or quantum fluctuations.
The contribution of attractor type saddle points into the partition function (\ref{eq:GRPartitionFunctionSmple})
is therefore significant. However, the key word here is {}``eventually''.
For any non-attractor solution it takes a time $t_{{\rm coll}}$ before the solution reaches its attractor
asymptotics.

Let us construct some initial state $|\Psi(t=t_i)\rangle$ of quantum
matter fields in a curved spacetime and gravitons. The amplitude $\langle\Psi(t_f)|\Psi(t_i)\rangle$ is
then defined by the path integral (\ref{eq:GRPartitionFunctionSmple}) calculated on
the closed Schwinger-Keldysh contour from $t=t_i$ to $t=t_f$ and
back. Then, if $t_{f}\ll t_{{\rm coll}}$, the corresponding attractor saddle point
does not give any noticeable contribution into the amplitude.%
\footnote{Of course, the time scale $t_{{\rm coll}}$ itself is a functional
of the initial state $|\Psi(t=t_i)\rangle$.%
} If it is necessary to know the evolution of the quantum state $|\Psi(t)\rangle$
at time scales $t\ll t_{{\rm coll}}$, we are forced to pay much more
attention to the type of saddle corresponding to general solutions of the Einstein equations.

Certainly, the Einstein equations are hard to solve, and it
is possible to find something like their general solution only in
physically simplified situations. As was first shown by Belinskii,
Khalatnikov and Lifshitz \cite{BKL}, \emph{asymptotically,} the general
solutions of the Einstein equations near the cosmological singularity
have the very same form for an almost arbitrary choice of the matter
content. This asymptotics in the synchronous frame\footnote{Often, it is impossible to choose the globally synchronous frame of reference due to the limitations set by the casuality. However, everywhere in the text we discuss the physics in a given casual patch.} is given by Kasner-like solution
\begin{equation}
ds^{2}=dt^{2}-\gamma_{\alpha\beta}(t,{\bf x})dx^{\alpha}dx^{\beta},
\label{eq:metricsynchro}
\end{equation}
\begin{equation}
\gamma_{\alpha\beta}(t,x)=t^{2p_{1}}{\bf l}_{\alpha}{\bf l}_{\beta}+t^{2p_{2}}{\bf m}_{\alpha}{\bf m}_{\beta}+t^{2p_{3}}{\bf n}_{\alpha}{\bf n}_{\beta}.
\label{eq:BKLgamma}
\end{equation}
 Both Kasner exponents $p_{1}$, $p_{2}$, $p_{3}$ and Kasner axis
vectors ${\bf l}_{\alpha}$, ${\bf m}_{\alpha}$ and ${\bf n}_{\alpha}$
are arbitrary functions of space coordinates. The Einstein equations
provide two constraints on the Kasner exponents
\begin{equation}
p_{1}+p_{2}+p_{3}=1,
\label{eq:KasnerConst1}
\end{equation}
and
\begin{equation}
p_{1}^{2}+p_{2}^{2}+p_{3}^{2}=1,
\label{eq:KasnerConst2}
\end{equation}
as well as three other constraints on arbitrary functions of space coordinates
present in (\ref{eq:BKLgamma}). Taking into account that the choice
of synchronous gauge
\begin{equation}
g_{00}=1,\; g_{0\alpha}=0
\label{eq:syncgauge}
\end{equation}
leaves the freedom to make three-dimensional space coordinate transformations,
one can easily see that the total number of arbitrary coordinate functions
in the Kasner-like solution (\ref{eq:metricsynchro}),(\ref{eq:BKLgamma})
is equal to $4$ as it should be expected for a general solution of
Einstein equations corresponding to an empty spacetime.

In the presence of the hydrodynamic matter Kasner solution (\ref{eq:metricsynchro}),(\ref{eq:BKLgamma})
describes asymptotic behavior of metrics near the singularity,
\footnote{Which corresponds everywhere below to the spacelike hypersurface $t=0$.%
} since components of energy-momentum tensor $T_{ik}$ grow slower
at $t\to0$ then the components of the Ricci tensor.%
\footnote{If there is a scalar field in the matter content \cite{BKLscalar}, BKL
solution (\ref{eq:BKLgamma}) remains general solution of the Einstein
equations with changed Kasner constraints (\ref{eq:KasnerConst1}),(\ref{eq:KasnerConst2}). %
} Higher order corrections to the Kasner solution (\ref{eq:metricsynchro}),(\ref{eq:BKLgamma}),
i.e., higher order terms in the expansion of $\gamma_{\alpha\beta}(t,x)$
over powers of $t$ play the role of perturbations which give rise
to the time dependence of Kasner exponents $p_{i}$ as well as Kasner
axis vectors ${\bf l}_{\alpha}$, ${\bf m}_{\alpha}$ and ${\bf n}_{\alpha}$
and to well-known BKL chaotic behavior. Therefore, the BKL solution
is simultaneously a \emph{universal attractor} for all solutions of
the Einstein equations possessing a spacelike singularity.
It means that \emph{no other saddle points contribute} into the amplitude
(\ref{eq:GRPartitionFunctionSmple}) in the vicinity of the cosmological
singularity.

In this essay, it will be first of all shown that in the presence
of a scalar field with potential $V(\phi)$ which is exponential and
unbounded from below, the general asymptotic solution of the Einstein
equations is different from the BKL solution and is quasi-isotropic
\cite{LK1} (while the BKL solution is essentially anisotropic). In
particular, we will choose potential of the form\footnote{The quasi-isotropic solution for such potentials was first found at the background level in \cite{Wesley}, where it was also shown that it is the attractor. The goal we pursue in this essay is to prove that the quasi-isotropic solution is also general and to understand how its instability develops   with the change of the form of the potential.}
\begin{equation}
V(\phi)=-|V_{0}|{\rm ch}\left(\lambda\phi\right).
\label{eq:ScalarFieldPotential}
\end{equation}
Scalar field potentials of this form appear in problems related to
gauged supergravity models \cite{GaugedSUGRA} and the ekpyrotic scenario
\cite{Ekpyrotic}. The cosmological singularity realized in such theory
is of the Anti de Sitter Big Crunch type. The physics in its vicinity
it is interesting by itself and even more so since
this type of singularity seems to be realized quite often on the string
theory landscape \cite{LindeLandscape}.

As in the case discussed in \cite{BKL}, it is convenient to perform
all calculations in the synchronous frame of reference where $g_{00}=1$,
$g_{0\alpha}=0$, $g_{\alpha\beta}=-\gamma_{\alpha\beta}$, $\alpha,\beta=1\ldots 3$,
i.e., the spacetime interval has the form
\begin{equation}
ds^{2}=dt^{2}-\gamma_{\alpha\beta}(t,x)dx^{\alpha}dx^{\beta}.
\label{metrics}
\end{equation}
Near the hypersurface $t=0$ which corresponds to the singularity,
the spatial metric components behave as
\begin{equation}
\gamma_{\alpha\beta}(t,{\bf x})=a_{\alpha\beta}({\bf x})t^{2q}+c_{\alpha\beta}({\bf x})t^{d}+b_{\alpha\beta}(x)t^{n}+\sum_{i,j}d_{\alpha\beta}^{(i,j)}(x)t^{f_{ij}}.
\label{metricseries}
\end{equation}
With the same precision, one has in the vicinity of singularity
\begin{equation}
\phi(t,{\bf x})=\psi(x)+\phi_{0}({\bf x}){\rm log}(t)+\phi_{1}(x)t^{f_{1}}+\phi_{2}(x)t^{f_{2}}+\cdots,
\label{phiseries}
\end{equation}
with dots corresponding to higher order terms of $\phi(t,{\bf x})$
expansion in powers of $t$. From the Einstein equations one finds%
\footnote{Due to the limitations of space we are unable to present the full derivation
of the solution here. It will be given in the forthcoming publication
\cite{PodolskyStarobinskyPrep}.%
} that the leading exponents in the expansions (\ref{metricseries})
and (\ref{phiseries}) are defined by the expressions%
\footnote{The indices of all matrices are lowered and raised by the tensor $a_{\alpha\beta}$,
for example, $b_{\alpha}^{\beta}=a^{\beta\gamma}b_{\gamma\alpha}$.%
}
\begin{eqnarray}
q & = & \frac{16\pi}{M_{P}^{2}\lambda^{2}},\,\, n=2,\,\, d=1-q,
\label{eq:qn1d1}
\end{eqnarray}
\begin{equation}
\psi(x)={\rm Const,}\,\,\phi_{0}(x)=\frac{2}{\lambda},\,\, f_{1}=1-3q,\,\, f_{2}=2-q,
\label{eq:phi0psi}
\end{equation}
\begin{equation}
c_{\alpha}^{\alpha}(x)=2\lambda\phi_{1}(x),\,\, c_{\alpha;\beta}^{\beta}(x)=\frac{1-2q}{1-3q}\frac{16\pi}{M_{P}^{2}}\phi_{0}\phi_{1,\alpha}(x),
\label{eq:c}
\end{equation}
\begin{equation}
\tilde{P}_{\alpha}^{\beta}(x)+(1-q)(qb_{\gamma}^{\gamma}(x)\delta_{\alpha}^{\beta}+(1+q)b_{\alpha}^{\beta}(x))=\frac{4\pi V_{0}}{M_{P}^{2}}e^{-\psi}\lambda\phi_{2}(x),
\label{eq:brecovery}
\end{equation}
\begin{equation}
-(1-q)b_{\alpha}^{\alpha}(x)
=\frac{32\pi}{M_{P}^{2}}(1-q)\phi_{0}\phi_{2}(x)-\frac{4\pi}{M_{P}^{2}}V_{0}e^{-\psi}\lambda\phi_{2}(x),
\label{eq:phi2recovery}
\end{equation}
where $\tilde{P}_{\alpha}^{\beta}(x)$ is the $3$-dimensional Ricci
tensor constructed from components of the tensor $a_{\alpha}^{\beta}(x)$
as components of metric tensor. Higher order terms in the expansions (\ref{metricseries})
and (\ref{phiseries}) can be selfconsistently calculated by using the
Einstein equations and the orthogonality condition
\begin{equation}
\gamma_{\alpha}^{\beta}\gamma_{\beta}^{\lambda}=\delta_{\alpha}^{\lambda}.
\label{eq:orthogonality}
\end{equation}
One can immediately find from Eq. (\ref{eq:orthogonality}) that the
higher order exponents in the metric (\ref{metricseries}) are defined
by
\begin{equation}
f_{ij}=i+2j-(3i+2j-2)q,
\label{eq:allmetricexponents}
\end{equation}
where $i,j\in\mathbb{N}$. The $n$ term in the metric expansion
corresponds to $i=0,\, j=1$ and $d$ term --- to $i=1,\,\, j=0$.
It is easy to see that there is no other exponents in the expansion
(\ref{metricseries}).

Let us examine the formulae (\ref{eq:qn1d1})-(\ref{eq:phi2recovery})
more closely and calculate the number of arbitrary functions present
in this solution. First of all, one can immediately see that the tensor
$a_{\alpha}^{\beta}(x)$ is not constrained by the Einstein equations.
It has $6$ components, and $3$ of them can be made to be equal to
$0$ by a three-dimensional coordinate transformation (the remnant gauge
freedom of the synchronous gauge (\ref{eq:syncgauge})). Since this
tensor is used for lowering and rising the indices and represents the
leading term in the expansion (\ref{metricseries}), we will identify
the term $a_{\alpha\beta}t^{2q}$ as a \emph{background} contribution
to $\gamma_{\alpha\beta}(t,x)$. Furthermore, we see from Eqs. (\ref{eq:brecovery}),(\ref{eq:phi2recovery})
that $b_{\alpha\beta}$ can be reconstructed from the known tensor
$a_{\alpha\beta}$.

The tensor $c_{\alpha\beta}$ contains three more arbitrary functions
of coordinates. Indeed,  it can be represented in the form
\begin{equation}
c_{\alpha}^{\beta}(x)=\frac{1}{3}c_{\gamma}^{\gamma}\delta_{\alpha}^{\beta}+Y_{;\alpha}^{\beta}+Y_{\alpha}^{;\beta}-\frac{2}{3}Y_{\gamma}^{\gamma}\delta_{\alpha}^{\beta}+c_{\alpha}^{{\rm (TT)}\beta}.
\label{eq:YorkDecomposition}
\end{equation}
From Eq. (\ref{eq:c}) one can see that its trace part defines the value
of $\phi_{2}(x)$ contributing to Eq. (\ref{phiseries}) and therefore
provides one arbitrary function. Then, three components of the vector
contribution $Y_{\alpha}(x)$ are fixed, and transverse traceless
part $c_{\alpha}^{{\rm (TT)}\beta}(x)$ provides remaining two arbitrary
functions. We also note that the $c_{\alpha\beta}$ term can be regarded
as the leading term \emph{perturbation} to the background contribution
into $\gamma_{\alpha\beta}$. In particular, it contains the contribution of scalar perturbations (related to the trace of the tensor $c_\alpha^\beta$) and tensor perturbations or gravitons (related to the transverse traceless part of the tensor $c_\alpha^\beta$).

The total number of arbitrary functions in the solution (\ref{metricseries}),(\ref{phiseries})
is therefore $6$, as one may expect for the general solution of the Einstein equations with a scalar field. By analysis similar to \cite{BKL}, one may show
\cite{Wesley,PodolskyStarobinskyPrep} that the contributions of other matter
fields into the overall energy-momentum tensor grow slower at $t\to0$
than the contribution of the scalar field. We conclude that the solution
(\ref{metricseries}), (\ref{phiseries}) is the general asymptotic
solution of Einstein equations (with arbitrary matter content) near
the cosmological singularity. Similarly to the BKL solution, the quasi-isotropic
solution is \emph{the universal attractor} for all solutions of the
Einstein equations with scalar field having the potential (\ref{eq:ScalarFieldPotential})
and arbitrary additional matter content which possess the time-like singularity.
Again, under considered conditions, no other saddle points
contribute into the amplitude (\ref{eq:GRPartitionFunctionSmple})
in the vicinity of the Big Crunch singularity.

It is instructive to understand how exactly the transition from the
quasi-isotropic regime (\ref{metricseries}),(\ref{phiseries}) near
the singularity to the BKL anisotropic regime (\ref{eq:BKLgamma})
happens. This transition can be achieved by changing the value of
$\lambda$while keeping $V_{0}$ fixed (or vise versa).

By construction, $2q<d=1-q$, i.e., the exponent
$a_{\alpha\beta}t^{2q}$ in the expansion (\ref{metricseries}) is
leading. With the increase of $q$, the value of $d$ decreases and
when $q$ reaches the critical value $q_{c}=1/3$, the contributions
$a_{\alpha\beta}(x)t^{2q}$ and $c_{\alpha\beta}(x)t^{d}$ into the expansion of the metric (\ref{metricseries})
become of the same order. Similarly, one can check that the values
of higher order exponents (\ref{eq:allmetricexponents}) decrease
with the increase of $q$. In particular, all exponents with different
$i$'s and similar $j$'s become of the same order of magnitude at
$q_{c}=1/3$. At $q>q_{c}=1/3$ the general asymptotic solution of
the Einstein equations near the singularity is given by Eq. (\ref{eq:BKLgamma})
instead of Eq. (\ref{metricseries}).

In fact, what we have just found is relevant for the quantum part
of the story, too, and in a sense is analogous to the \emph{spontaneous
symmetry breaking} phenomenon in QFTs. Indeed, let us take the theory
with a scalar field
\begin{equation}
\mathcal{L}=\frac{1}{2}\partial_{i}\Phi\partial^{i}\Phi-\frac{\lambda}{4}(\Phi^{2}-v)^{2},
\label{eq:SpontSymmBreak}
\end{equation}
set $\Phi(x,0)=0$ as an initial condition and continuously change the value of the parameter $v$. At $v>0$
the solution $\Phi(t,x)=0$ of the classical equations of motion is
perturbatively stable and corresponds to the true vacuum of the theory
at the quantum level. At $\nu<0$ the same solution becomes classically
unstable, and $\Phi(t,x)$ reaches the {}``true'' vacuum value $\Phi=\pm\sqrt{v}$
during the time $t\sim\frac{1}{\lambda\sqrt{v}}{\rm log}\frac{1}{\lambda}$
(with the VEV of the operator $\hat{\Phi}$ having similar behavior
at the quantum level). Similar situation is realized in our case.

At $q<q_{c}=1/3$ the quasi-isotropic solution (\ref{metricseries}),(\ref{phiseries})
is the general solution of the Einstein equations; it is perturbatively
stable by construction (without any limitations on the weakness of
the perturbations). At $q>q_{c}$ the quasi-isotropic solution becomes
perturbatively unstable (perturbations defined by $c_{\alpha\beta}$
and higher order terms grow faster than the background term $a_{\alpha\beta}$
at $t\to0$).

Vise versa, at $q>q_c=1/3$ the BKL anisotropic solution if the Einstein equations is general in
the vicinity of the cosmological singularity. It is stable by construction with respect to arbitrary perturbations and the stability is lost at $q<q_c$.

This analysis remains valid for the quantum situation\footnote{One important comment regarding the quantization should be made. The quantum theory of the scalar field  with the potential (\ref{eq:ScalarFieldPotential}) is tachyonically unstable and has neither well-defined asymptotic $|{\rm out}\rangle$ states, nor $\langle {\rm out}|{\rm in} \rangle$ S-matrix. However, the Schwinger-Keldysh $\langle {\rm in}|{\rm in} \rangle$ S-matrix is defined, and it is possible to make sense of the corresponding time-dependent theory \cite{PodolskyStarobinskyPrep}.} since the canonical phase space is in one-to-one correspondence with the space of solutions of classical field equations \cite{CWZ}, and both quasi-isotropic and BKL solutions are (a) general and (b) universal attractors for other solutions of the Einstein equations in the vicinity of the time-like singularity.

The transition from the regime realized at $q<1/3$ to the regime
$q>1/3$ probably corresponds in the quantum level to the condensation
of gravitational perturbations. Indeed, one can interpret the higher order contributions in the expansion (\ref{metricseries}) as terms corresponding to the \emph{interaction} between gravitational degrees of freedom as well as higher order nonlinearities in the background. Our conclusion is based on the fact that at $q=q_c$ the spectrum of the exponents in the expansion (\ref{metricseries}) becomes infinitely dense.
It is also possible to show that the point of the {}``phase transition''
$q_{c}=1/3$ corresponds at the classical level to the situation when
the choice of globally synchronous frame of reference is impossible
near the singularity \cite{PodolskyStarobinskyPrep}.

Let us summarize what have been found in the present essay. We have
shown that in the presence of the scalar field with exponential potential unbounded
from below, the general asymptotic solution of the Einstein equations
near the cosmological singularity has quasi-isotropic behavior instead
of anisotropic found by \cite{BKL}. We have argued that at the
quantum level there should exist a phase transition between the quasi-isotropic
and anisotropic phases, governed by the strength of the scalar field
potential and interpreted this phase transition as the condensation
of gravitational perturbations.

\subsection*{Acknowledgements}

I am thankful to A.A. Starobinsky and D. Wesley for the discussions and to K. Enqvist for making helpful comments. While conducting this work, I was supported by Marie Curie Research
training network HPRN-CT-2006-035863.

\end{document}